\documentclass[twocolumn,showpacs,amsmath,amssymb,prl]{revtex4}

\usepackage{graphicx}

\newcommand{\be}{\begin{equation}}
\newcommand{\ee}{\end{equation}}
\newcommand{\bfig}{\begin{figure}}
\newcommand{\efig}{\end{figure}}

\begin{document}
\title{Quantum oscillations in topological superconductor candidate Cu$_{0.25}$Bi$_2$Se$_3$
}
\author{Ben J. Lawson$^1$, Y. S. Hor$^2$, Lu Li$^1$ 
}
\affiliation{
$^1$Department of Physics, University of Michigan, Ann Arbor, MI  48109\\
$^2$Department of Physics, Missouri University of Science and Technology, Rolla, MO 65409
}

\date{\today}
\begin{abstract}
 Quantum oscillations are generally studied to resolve the electronic structure of topological insulators. In Cu$_{0.25}$Bi$_2$Se$_3$, the prime candidate of topological superconductors, quantum oscillations are still not observed in magnetotransport measurement. However, using torque magnetometry,  quantum oscillations (the de Hass - van Alphen effect) were observed in Cu$_{0.25}$Bi$_2$Se$_3$ . The doping of Cu in Bi$_2$Se$_3$ increases the carrier density and the effective mass without increasing the scattering rate or decreasing the mean free path.  In addition, the Fermi velocity remains the same in Cu$_{0.25}$Bi$_2$Se$_3$ as that in Bi$_2$Se$_3$. Our results imply that the insertion of Cu does not change the band structure of Bi$_2$Se$_3$.
\end{abstract}


\maketitle                   
Topological insulators and topological superconductors are new families of materials with novel electronic states \cite{K1, SCZ1, Mol07, H1, Shen1, H5, Hor_L10, HK_RMP10, QZ_RMP11}. Bi$_2$Se$_3$ is a topological insulator (TI) that has attracted special interest since Cu-intercalation between quintuple layers of Bi$_2$Se$_3$ can induce superconductivity below 3.8 K~\cite{Hor_L10}. Unconventional superconductivity arising from these topological materials may support Majorana fermions and provide a platform for topological quantum computation \cite{FK_Majorana, Wil_NP09, M_N10}. It is proposed that this electron-doped topological insulator Cu$_x$Bi$_2$Se$_3$ is a topological superconductor (TSC)~\cite{FB_L10}. Instead of a bulk insulating gap and a surface Dirac state, a TSC has a full pairing gap in the bulk and a topologically protected gapless Andreev bound state consisting of Majorana fermions on the surface. Specific-heat measurement later elucidated that Cu$_x$Bi$_2$Se$_3$ is a bulk strong-coupling superconductor with a full pairing gap~\cite{MKR_L11}. Furthermore, from a point-contact spectroscopy study, the Ando group has discovered that this Cu$_x$Bi$_2$Se$_3$ exhibits a surface Andreev bound state~\cite{Ando_point}. All these analyses of possible superconducting gap functions have allowed the conclusion that Cu$_x$Bi$_2$Se$_3$ is indeed a TSC. This is a big step towards the goal of identifying the new topological phase of matter. However, the observation of quantum oscillations, which are generally studied to resolve the electronic structure of topological materials, is still missing in Cu$_x$Bi$_2$Se$_3$,  the leading candidate of TSC. For a TSC, as well as the TIs such as Bi$_2$Se$_3$ and Bi$_2$Te$_3$, it is necessary to observe quantum oscillations in order to resolve Landau level quantization and yield direct measurement of the bulk and surface electronic states \cite{Ong_interference, Ando_BiSb, Ando_BiSe, Ong_BiTe, Analytis_NP09}.  As a result,  an exact measurement of the effective mass and the scattering rate of this TSC material is still controversial \cite{Wray_NP10, MKR_L11, MKR2, Wray_PRB11}. 

We solved the problem by measuring the magnetic torque of a single crystalline  Cu$_{0.25}$Bi$_2$Se$_3$. Quantum oscillations measured in torque magnetometry, known as  the de Hass - van Alphen effect, are observed in the high field state of the fully superconducting  Cu$_{0.25}$Bi$_2$Se$_3$. Compared with the oscillation pattern in Bi$_2$Se$_3$, the Fermi surface of Cu$_{0.25}$Bi$_2$Se$_3$ becomes larger and stays ellipsoidal. We resolve the effective mass and the scattering rate in the crystal ab plane. The effective mass of Cu$_{0.25}$Bi$_2$Se$_3$ is found to be enhanced to  0.19 $m_e$ (free electron mass), compared with 0.14 $m_e$ from the bulk state of Bi$_2$Se$_3$. Moreover, the scattering rate and the Fermi velocity stay the same in the Cu$_{0.25}$Bi$_2$Se$_3$, as those in Bi$_2$Se$_3$.

\begin{figure}[t]
\includegraphics[width=3.2 in ]{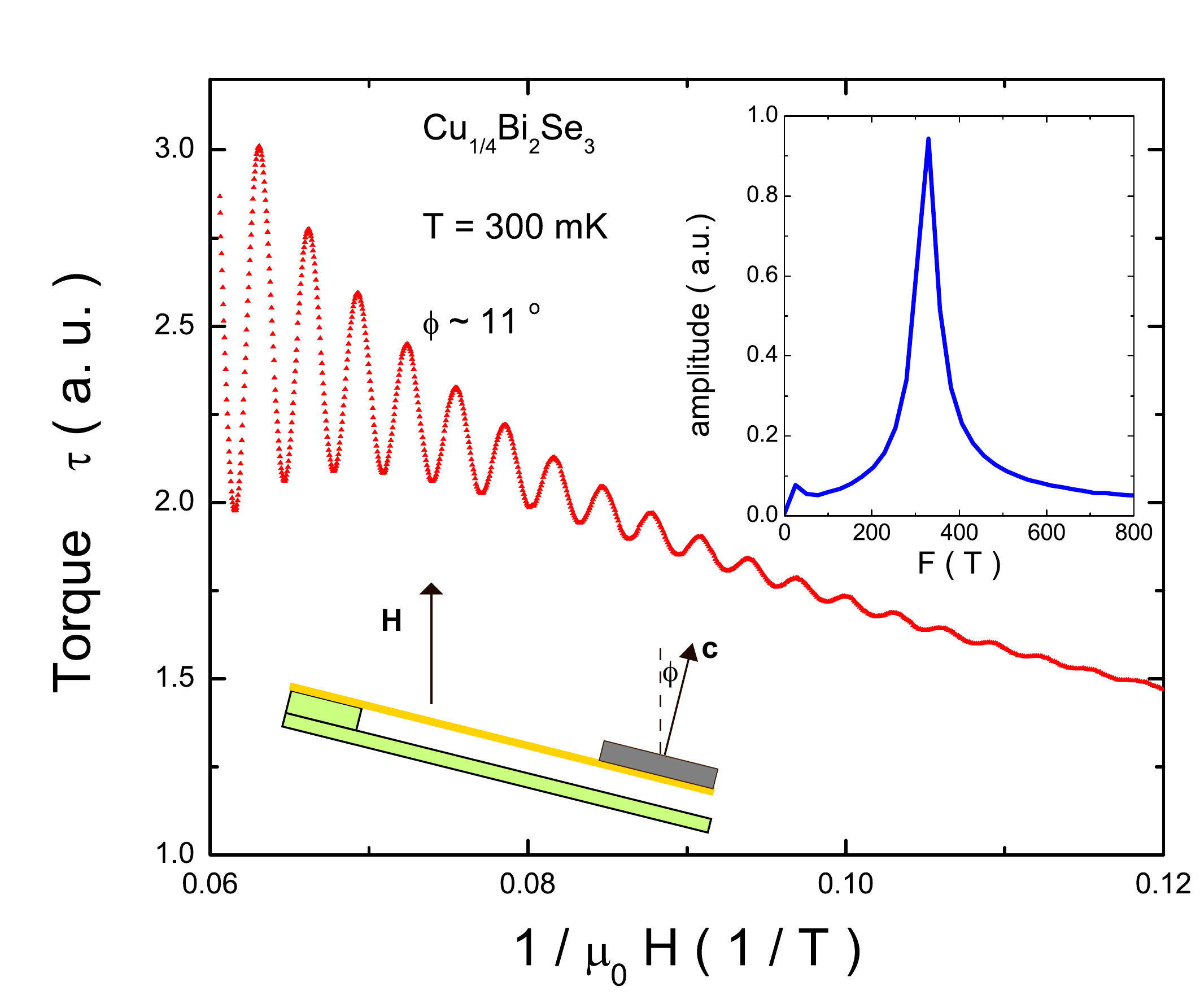}
\caption{\label{figTorque} (color online)
Quantum oscillation observed by torque magnetometry. Magnetic torque $\tau$ is plotted as a function of 1/$\mu_0H$. The lower left panel shows the sketch of the measurement setup, where the magnetic field is applied to the sample with a tilt angle $\phi$ relative to the crystalline $c$ axis. The Fast Fourier Transform (FFT) plot of the torque signal is shown in the upper right panel. To generate a FFT plot from the $\tau - 1/\mu_0H$ data, the polynomial background is subtracted before the FFT.
}
\end{figure}

Cu$_{0.25}$Bi$_2$Se$_3$ single crystals were grown by slow cooling of melted stoichiometric mixtures of high purity elements Bi (99.999 \%), Cu (99.99 \%) and Se (99.999 \%) from 850$^\mathrm{o}$C to 620 $^\mathrm{o}$C in a sealed evacuated quartz tube. The crystals were then quenched  at 620 $^\mathrm{o}$C in cold water, resulting in easily cleaved crystals with shiny mirror like surfaces. However, the silvery surfaces turn golden after one day of exposure to air. The chemical formula of Cu$_{0.25}$Bi$_2$Se$_3$ was determined according to the mole ratio of the starting elements used in the crystal growth.

Torque magnetometry was applied to measure the anisotropy of magnetic susceptibilities of samples under external magnetic fields \cite{PRLanisotropyBSCCO,NatureOscillationSebastian,LiNatPhys07}. With the tilted magnetic field $\bf H$ confined to the $x$-$z$ plane, and $M$ in the same plane, the torque is $\vec{\tau} = \vec{M} \times \vec{H} = (M_zH_x - M_xH_z){\bf\hat{y}}$. In the normal state of samples, where the magnetization is strictly linear in $H$, it is convenient to express
the observed torque as a function of the susceptibility difference as follows,
\begin{align}
\tau  &=  \chi_zH_zH_x-\chi_xH_xH_z  \notag \\
        &= (\chi_z-\chi_x)H_xH_z   =  \Delta\chi H^2\sin\phi\cos\phi
\end{align}
\label{chilin}

where $\phi$ is the tilt angle of $\vec{H}$ away from $\bf\hat{z}$, and $\Delta\chi = \chi_z-\chi_x$ is the magnetic susceptibility anisotropy.  In this letter, the crystalline $\bf c$ axis of the single crystal Cu$_{0.25}$Bi$_2$Se$_3$ is parallel to the $\bf\hat{z}$ axis.
 
In our experimental setup, a Cu$_{0.25}$Bi$_2$Se$_3$ single crystal is glued to the tip of a thin metal cantilever, as shown in the lower panel of Fig. \ref{figTorque}. The magnetic torque $\tau$ is measured capacitively. Torque $\tau$ is plotted as a function of the inverse of the applied magnetic field $\mu_0H$, displayed in Fig. \ref{figTorque} at temperature $T$  = 300 mK and $\phi \sim 11^{\circ}$.  Oscillatory pattern in $\tau$ is periodic in $1/\mu_0H$, reflecting the quantization of the landau levels. For metals, the oscillating frequency $F_S$ is determined by the cross section area $A$ of the Fermi surface, as following,
\begin{equation}
F_S = \frac{\hbar}{2\pi e} A.
\label{Fs}
\end{equation}
In the upper right panel of Fig. \ref{figTorque}, the Fast Fourier Transformation (FFT) of the $\tau$ shows a single peak in the amplitude spectrum with a corresponding $F_S\sim$ 325 T.

\begin{figure}[h]
\includegraphics[width=3.2 in]{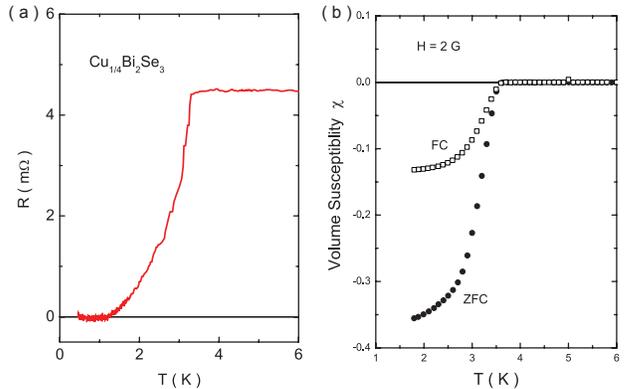}
\caption{\label{figRTChiT} (color online)
The temperature $T$ dependence of the sample resistance $R$. (Panel a) and magnetic susceptibility $\chi$ (Panel b).  (Panel a)Zero resistance is observed below 1.2 K. (Panel b) Volume magnetic susceptibility $\chi$ is measured at field $H$ = 2 G in the crystal $ab$ plane, in both the  Zero-Field-Cooled (ZFC) and the Field-cooled (FC) conditions. From the ZFC curve,  the nominal superconducting fraction is found to be around 35 \%, consistent with the fraction found in earlier reports \cite{MKR_L11, MKR2}.
}
\end{figure}

The dHvA effect is also observed in the normal state of our Cu$_{0.25}$Bi$_2$Se$_3$ samples. One may suspect an extreme picture that phase separation within the samples forms two different domains: a Cu-rich region which is superconducting and another Cu-deficient region where simply the undoped Bi$_2$Se$_3$  produces a quantum oscillation signal. The first challenge is to figure out whether the oscillatory pattern comes from the Cu-doped superconducting phase or from the undoped Bi$_2$Se$_3$ phase. To answer this question, we first present the evidence of zero-resistance and the Meissner effect that are consistent with earlier reports from other groups. Then, we compare the dHvA effect of Cu$_{0.25}$Bi$_2$Se$_3$ and Bi$_2$Se$_3$. We found that the Cu doping changes the quantum oscillation pattern, by increasing the carrier density as well as the effective mass, but it does not change the Fermi velocity or the scattering rate. We conclude that the quantum oscillation is indeed coming from the Cu-doped phase.

The four-probe resistance $R$ of the sample is shown as a function of $T$ in Fig.\ref{figRTChiT}(a). As $T$ decreases, $R$ starts to drop at 3.3 K, decreases quickly at $T$ = 3 K, and becomes zero below $T$ = 1.2 K. The low field magnetic susceptibility was also measured with a Quantum Design MPMS magnetometer. Fig.\ref{figRTChiT}(b) shows the volume magnetic susceptibility $\chi = M/\mu_0H$ at 1.8 K $\leq T \leq$ 6 K at $H$ = 2 G in-plane magnetic field. In the Field-cooled (FC) run, the sample was cooled down to 1.8 K under the 2 G external field and $\chi$ was measured during the warm up. By contrast, in the Zero-Field-cooled (ZFC) run, the sample was cooled down at zero external field.  In both runs, rapid decrease of $\chi$ was observed at $T \sim$ 3.5 K. Moreover, at the lowest achieved $T\sim$ 1.8 K in the MPMS system, $\chi \sim$ -0.35 suggested  a 35\% superconducting volume, which is comparable to other reports \cite{MKR_L11, MKR2}. This considerably large superconducting fraction suggests that our Cu$_{0.25}$Bi$_2$Se$_3$ sample is indeed high quality single phased crystal. Moreover, the observation of the zero resistance implies that a large scale phase separation does not occur in our single crystals.

\begin{figure}[h]
\includegraphics[width=3.2 in]{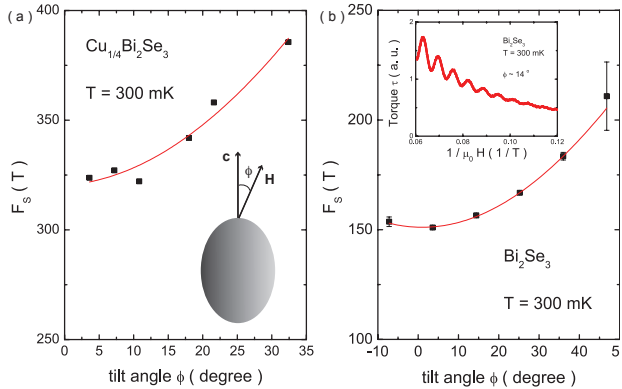}
\caption{\label{figAngle} (color online)
Angular dependence of the oscillating period $F_S$ is compared between (Panel a) Cu$_{0.25}$Bi$_2$Se$_3$ and (Panel b) Bi$_2$Se$_3$, Solid lines are the fits based on a single ellipsoidal Fermi surface. An example of the oscillating magnetic torque is shown in the inset of Panel b for Bi$_2$Se$_3$.
}
\end{figure}

Further evidence shows the effect of Cu doping by comparing the dHvA effect of the superconductor Cu$_{0.25}$Bi$_2$Se$_3$ and the topological insulator Bi$_2$Se$_3$. Fig. \ref{figAngle} shows the oscillating period $F_S$ at low tilt angles measured in Cu$_{0.25}$Bi$_2$Se$_3$ (Panel a) and Bi$_2$Se$_3$ (Panel b).  First of all, $F_S$ for Cu$_{0.25}$Bi$_2$Se$_3$ is much larger than that of the undoped sample, which suggests the Cu doping indeed adds carriers into the electronic state. For both samples, the oscillation from the bulk state shows only one oscillation frequency, suggesting a single ellipsoidal Fermi pocket in both samples. Compared with Bi$_2$Se$_3$, Cu$_{0.25}$Bi$_2$Se$_3$ shows the following three features.

{\it 1. Larger Fermi pocket.}  
The angular dependence of $F_s$ provides an estimate of the Fermi pocket size. The $F_s - \phi$ curve n  Cu$_{0.25}$Bi$_2$Se$_3$ is consistent with an ellipsoidal Fermi surface with  the cross section $A_{xz}$ about 4.02 nm$^{-2}$. Base on these results and Eq. \ref{Fs}, we obtain $k^x_F = k^y_F =$ 0.97 nm$^{-1}$, and $k^z_F =$ 1.3 nm$^{-1}$. Thus, the bulk carrier density $n = \frac{1}{3\pi^2}k^x_Fk^y_Fk^z_F$ is 4.3 $\times$ 10$^{19}$ cm$^{-3}$. This carrier density is similar to the estimate based on the Hall effect \cite{MKR_L11} and ARPES \cite{Wray_NP10}. For the control sample Bi$_2$Se$_3$, the sinusoidal fitting gives $k^x_F = k^y_F =$ 0.69 nm$^{-1}$ and  $k^z_F =$ 1.2 nm$^{-1}$. The bulk carrier density $n$ is calculated to be 1.8$\times$ 10$^{19}$ cm$^{-3}$. This number is larger than those generally seen in results based on quantum oscillations \cite{Ando_BiSe}, which suggests that there are quite a number of carriers caused by defects.

We note that not only the carrier density is increased in the Cu doped topological superconductors, but also the electronic state is closer to a three dimensional sphere. The aspect ratio $k^z_F/k^x_F$ of  Bi$_2$Se$_3$ is about 1.6, consistent with ref. ~\cite{Ando_BiSe}. By contrast,  $k^z_F/k^x_F$ is about 1.3 in Cu$_{0.25}$Bi$_2$Se$_3$. The doping of Cu changes the relative ratio of Fermi ellipsoid and makes it closer to a sphere.

\begin{figure}[h]
\includegraphics[width=3.2 in]{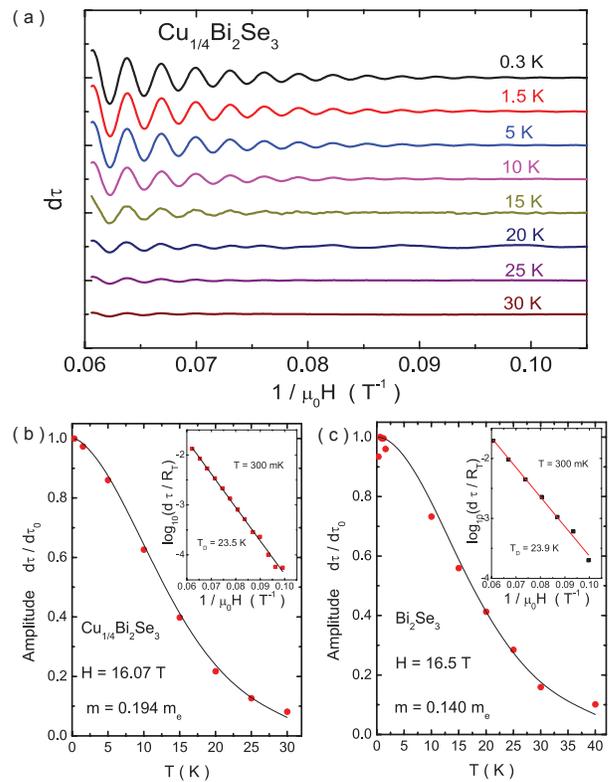}
\caption{\label{massDingle} (color online)
Temperature dependence of oscillating magnetic torque. (Panel a) Magnetic torque after subtracting a polynomial background $d\tau$ is plotted as a function of $1/H$ with $\phi \sim$ 4$^{\circ}$ at selected $T$ between 0.3 K and 30 K. Curves at different $T$ are displaced for clarity. (Panel b) In Cu$_{0.25}$Bi$_2$Se$_3$, temperature dependence of an oscillating amplitude $d\tau$ yield the effective mass $m = 0.194 m_e$. The dingle plot is shown in the inset, giving the Dingle temperature $T_D$ = 23.5 K. (Panel c) Comparison plot in Bi$_2$Se$_3$. The $T$ dependence of $d\tau$ at H = 16.5 T yields $m = 0.140 m_e$, and the Dingle plot at 300 mK shows $T_D$ = 23.9 K.
}
\end{figure}


{\it 2. Slightly increased effective mass.} We now focus on the temperature dependence and magnetic field dependence of the oscillating amplitude of magnetic torque as the magnetic field is along the $\bf{c}$-axis. The oscillating torque $d\tau$ is obtained by subtracting a monotonic background from the raw torque data.  In metals, the first harmonic of the oscillating magnetic torque is well described based on the LK formula ~\cite{Shoenberg}, and $d\tau$ is proportional to the thermal damping factor $R_T$ and the Dingle damping factor $R_D$, as follows,

\begin{align}
R_T &= \alpha Tm^*/B\sinh(\alpha Tm^*/B) \notag \\
R_D & = \exp(-\alpha T_Dm/B)
\label{LK}
\end{align}
where the effective mass $m = m^*m_e$, the Dingle temperature $T_D = \hbar/2\pi k_B \tau_S$. $\tau_S$ is the scattering rate , $m_e$ is the free electron mass, and  $\alpha = 2\pi^2 k_B m_e/e \hbar \sim $14.69 T/K \cite{Shoenberg}.

Fig. \ref{massDingle}(a) displays the oscillating amplitude $d\tau$ vs. $1/\mu_0H$ at selected $T$ between 300 mK and 30 K.  Offsets are added to these curves for clarity. The tilt angle $\phi \sim$ 4$^{\circ}$. All the following fitting results are for $H$ closely along the $\bf{c}$ -axis.

First, under a fixed field, the $T$ dependence of normalized $d\tau$ is determined by the effective mass $m^*m_e$. Fig. \ref{massDingle}(b) shows the $d\tau$ vs. $T$ at $H =$ 16.07 T.  The oscillating amplitude $d\tau$ is normalized by $d\tau_0$, the amplitude at the lowest $T = $ 300 mK.  Using the thermal damping $R_T$ formula in Eq. \ref{LK}, a fit  shown in a line yields the effective mass $m$ = 0.194 $m_e$.

By comparison, the same analysis using the oscillating torque $d\tau$ at  $H$ = 16.5 T in Bi$_2$Se$_3$ is shown in Fig. \ref{massDingle} (c). The $R_T$ fitting yields $m$  = 0.140 $m_e$, which agrees with earlier reports~\cite{Ando_BiSe}.  The comparison shows that  the effective mass increases gently in the topological superconductor candidate Cu$_{0.25}$Bi$_2$Se$_3$ . However, our result does not agree with the large mass-enhancement suggested by heat capacity measurements \cite{MKR_L11}.

{\it 3. Similar Fermi velocity and scattering rate.} Besides adding more carriers and increasing the effective mass, the Cu doping does not change the band structure of the parent compound Bi$_2$Se$_3$. This can be shown as we calculate the Fermi velocity $v_F = \hbar k_F/m$. Based on the effective mass values and $k_F^x$, we obtain $v_F$ = 5.8 $\times$ 10$^6$ m/s for Cu$_{0.25}$Bi$_2$Se$_3$, and $v_F$ = 5.7 $\times$ 10$^6$ m/s for Bi$_2$Se$_3$. Since $v_F$ determines the slope of the $E-k$ dispersion at the Fermi surface, the relatively similar values of $v_F$ in both samples suggested the added electrons by Cu doping enters the same mobile bulk band in Bi$_2$Se$_3$.

Further analysis of $H$ dependence of the oscillating amplitude gives a similar scattering rate in both Cu$_{0.25}$Bi$_2$Se$_3$ and Bi$_2$Se$_3$. The inset of Fig. \ref{massDingle}(b) displays the ``Dingle plot'' at 300 mK in Cu$_{0.25}$Bi$_2$Se$_3$. The fit yields the Dingle temperature $T_D$ = 23.5 K, which gives the scattering rate $\tau_S$ = 5.2 $\times$ 10$^{-14}$ s. Thus, the mean free path $l = v_F \tau_S$ = 30 nm. In Bi$_2$Se$_3$, the same fitting shown as the Dingle plot in Fig. \ref{massDingle}(c) yields $T_D$ = 23.9 K, $\tau_S$ = 5.1$\times$ 10$^{-14}$ s, and $l$ = 29 nm.

\begin{table}[h]
\caption{Parameters in Cu$_{0.25}$Bi$_2$Se$_3$ and Bi$_2$Se$_3$}
\centering
\begin{ruledtabular}
\begin{tabular}{c c c}

                      			& Cu$_{0.25}$Bi$_2$Se$_3$			&Bi$_2$Se$_3$  \\
\hline
$F_S $  				&  325 T							&150 T			\\
$n$					&  4.3$\times$ 10$^{19}$ cm$^{-3}$	&1.8$\times$ 10$^{19}$ cm$^{-3}$	\\
$k_F^x$				& 0.97 nm$^{-1}$					& 0.69 nm$^{-1}$ \\
$k_F^z/k_F^x$			&  1.3							& 1.6				\\
$m$					& 0.194 $m_e$						& 0.140 $ m_e$ \\
$v_F$				& 5.8 $\times$ 10$^6$ m/s			& 5.7 $\times$ 10$^6$ m/s \\
$\tau_S$				& 5.2 $\times$ 10$^{-14}$ s			& 5.1 $\times$ 10$^{-14}$ s \\
$l$					& 30 nm							& 29 nm \\

\end{tabular}
\end{ruledtabular}
\label{parameter}
\end{table}

{\textit Discussion}

Table \ref{parameter} summarizes the analysis results of the dHvA effect in Cu$_{0.25}$Bi$_2$Se$_3$ and Bi$_2$Se$_3$.  The doping of Cu increases the carrier density and the effective mass, but does not affect several key band structure parameters especially $v_F$.  As $v_F$ determines the slope of the energy dispersion at the chemical potential, the unchanged $v_F$ implies that the added carriers by Cu doping go into the same conductive band in the undoped Bi$_2$Se$_3$.  Note that $k_F^x$ increases by more then 40 \% in a Cu$_{0.25}$Bi$_2$Se$_3$, which should leads to a large increase in  $v_F$ for a quadratic band. Our observed unchanged $v_F$ suggests a rather linear Dirac-like band. 

The ratio of $T_D$ to $T_C$ also reflects the relationship between the scattering ratio $\lambda = 1/ \tau_S = 2 \pi k_B T_D/\hbar$ and the superconducting gap. Our result shows that $T_D =$ 23.5 K is much larger than $T_C = $ 3.5 K. Assuming the BCS gap $2\Delta = 3.5 k_B T_C$,  we find the dimensionless ratio $k_B T_D/ \Delta  \sim$ 3.8. Therefore, the superconductivity in Cu$_{0.25}$Bi$_2$Se$_3$ occurs in the dirty limit.  Further,  if the $\Delta/T_C$ ratio follows the BCS theory, the Pippard length $\xi_P = \hbar v_F/\pi\Delta$ is  2.3 $\mu$m, which is about two orders of magnitude larger than the mean free path $l$. As a result, the coherence length is mainly determined by the mean free path and the impurity scattering. We note that the $ab-$ plane coherence length determined by the upper critical field is 13.9 nm~\cite{MKR_L11}, similar to the mean free path.

Topological odd-parity superconductivity is proposed \cite{FB_L10} to exist in  Cu$_{0.25}$Bi$_2$Se$_3$. Experimental supports include a recent observed zero-bias conductance peak in the point-contact spectrum \cite{Ando_point}. For a general pairing based on the BCS theory, regular s-wave pairing is protected from the nonmagnetic impurities \cite{anderson}, whereas odd-parity pairing is easily destroyed by the impurity scattering \cite{disorder,sr2ruo4}. Our observation of the unchanged $v_F$ implies that the basic band parameter remains the same as Cu is doped to the topological insulator Bi$_2$Se$_3$. Reference \cite{MF_SO} suggests the strong spin-orbit locking in this band structure makes the odd-parity pairing robust against impurity scattering.

In conclusion, quantum oscillation is observed in the magnetization of  Cu$_{0.25}$Bi$_2$Se$_3$. Based on the quantum oscillating pattern, only one Fermi pocket exists in Cu$_{0.25}$Bi$_2$Se$_3$. The doping of Cu to the topological insulator Bi$_2$Se$_3$ increases the carrier density and the effective mass without damaging  the scattering rate and the mean free path. The Fermi velocity stays the same after the Cu doping, which implies the band structure is not affected by the insertion of Cu. Therefore, the analysis of the quantum oscillations has shown another important piece of evidence that Cu$_{0.25}$Bi$_2$Se$_3$ is likely a topological superconductor. Further studies of quantum oscillations in other Cu doping levels will reveal how the Cu doping tunes the band structure and the fate of topological superconductivity.

The authors gratefully acknowledge helpful discussions with Liang Fu. The work is supported by the University of Michigan (LL, BJL) and MST (YSH). The high-field experiments were performed at the National High Magnetic Field Laboratory, which is supported by NSF Cooperative Agreement No. DMR-084173, by the State of Florida, and by the DOE.

\end{document}